\documentclass[showpacs,preprintnumbers,amsmath,amssymb]{revtex4}

\usepackage{graphicx}
\usepackage{dcolumn}
\usepackage{bm}
 
\def\mbi#1{\mbox{\bfseries\itshape #1}} 

\begin{document}

\preprint{APS/123-QED}

\title{New Constraints on the Primordial Magnetic Field}

\author{Dai G. Yamazaki$^{1}$}
 \email{yamazaki@asiaa.sinica.edu.tw}
\author{Kiyotomo Ichiki$^{2}$}%
\author{Toshitaka Kajino$^{3,4}$}%
\author{Grant. J. Mathews$^{5}$}%
\affiliation{%
$^{1}$Institute of Astronomy and Astrophysics, Academia Sinica,
11F of Astronomy-Mathematics Building,
National Taiwan University.No.1, Roosevelt Rd, Sec. 4 Taipei 10617, Taiwan, R.O.C
}%
\affiliation{%
$^{2}$Department of Physics and Astrophysics, Nagoya University, Nagoya 464-8602, Japan
}%
\affiliation{%
$^{3}$
National Astronomical Observatory of Japan, Mitaka, Tokyo 181-8588, Japan
}%
\affiliation{%
$^{4}$
Department of Astronomy, Graduate School of Science, The University of Tokyo,
Hongo 7-3-1, Bunkyo-ku, Tokyo 113-0033, Japan
}%
\affiliation{%
$^{5}$Center for Astrophysics,
Department of Physics, University of Notre Dame, Notre Dame, IN 46556, U.S.A.
}%

\date{\today}

\begin{abstract}
We present the newest statistical and numerical analysis of the matter and cosmic microwave background power spectrum with effects of the primordial magnetic field (PMF) included.
New limits to the PMF strength and power spectral index are obtained based upon the accumulated data for both the matter and CMB power spectra on small angular scales.
We find that a maximum develops in the probability distribution for a magnitude of the PMF of 
$|B_\lambda| = 0.85 \pm 1.25(\pm 1\sigma)$ nG on a comoving scale of at 1 Mpc, corresponding to upper limits of 
$\mathbf{< 2.10 nG} ( 68\% \mathrm{CL})$
and 
$\mathbf{< 2.98 nG} ( 95\% \mathrm{CL})$. 
While for the power spectral index we find  
$n_\mathrm{B}= -2.37^{+0.88}_{-0.73}(\pm 1\sigma)$, 
corresponding to upper limits of
$ \mathbf{< -1.19} ( 68\% \mathrm{CL})$
and
$ \mathbf{< -0.25} ( 95\% \mathrm{CL})$. 
This result provides new constraints on models for  magnetic field generation and the physics of  the early universe.
We conclude that future observational programs for the CMB and matter power spectrum will likely provide  not only upper limits but also lower limits to the PMF parameters.  
\end{abstract}

\maketitle
\subsection{Introduction}
Magnetic fields are everywhere in nature, and they affect many physical processes over a wide range of scales in both space and time.
Indeed, stochastic magnetic fields have been observed to exist even on the scale of
galaxy clusters\cite{Wolfe:1992ab,Clarke:2000bz,Xu:2005rb} and their origin is  a recent focus in cosmology\cite{Turner:1987bw,Ratra:1991bn,Vachaspati:1991nm,Kibble:1995aa,Ahonen:1997wh,Joyce:1997uy,Barrow:1997mj,Durrer:1999bk,Bamba:2004cu,Takahashi:2005nd,Hanayama:2005hd,ichiki:2006sc}.
Although cluster magnetic fields may be explained by a dynamo mechanism, for example, occurring in AGN jets \cite{Xu:2009if} or via the Weibel instability \cite{Fujita:2006}, its existence on large scales allows for the possibility that a primordial magnetic field (PMF) could also be present.

Such a PMF could have influenced a variety of phenomena in the early universe
such as the cosmic microwave background
\cite{Jedamzik:1999bm,Seshadri:2000ky,Mack:2001gc,Subramanian:2002nh,Subramanian:2003sh,Lewis:2004ef,Yamazaki:2004vq,Yamazaki:2006bq,Yamazaki:2007oc,Giovannini:2008aa,Kojima:2008rf,Paoletti:2008ck,Caprini:2009vk,Kojima:2009ms},
non-Gaussianity and low multipole anomalies of the CMB\cite{Bernui:2008ve,Seshadri:2009sy,Caprini:2009vk},
the formation of large-scale structure (LSS)
\cite{
Tsagas:1999ft,
Giovannini:2004aw,
Tashiro:2005hc,
Tashiro:2005ua,
Yamazaki:2006mi,
Sethi:2008eq,
Yamazaki:2008bb}, and the gravitational wave background\cite{Caprini:2001nb}.
Effects of a PMF on the BB polarization mode of the CMB have also been  studied by several authors
\cite{Mack:2001gc,Seshadri:2000ky,Lewis:2004ef,Kosowsky:2004zh,Yamazaki:2007oc,Kristiansen:2008tx,Giovannini:2008zv,Kahniashvili:2008hx}
and these effects have proven to be as important as 
the other physical processes which strongly affect the BB polarization through Lorentz symmetry violation \cite{Komatsu:2008hk,Kostelecky:2009zp,Kamionkowski:2008fp}.

The purpose of this paper is to present the newest results of constraints on cosmological parameters by taking account of the effects of the PMF inferred from the Markov Chain Monte Carlo (MCMC) analysis of the observed CMB anisotropies and matter power spectrum.
We show that presently existing and future precise observations of the CMB anisotropies have the potential to place stringent constraints on the PMF parameters.
Indeed, existing data may already provide some evidence that a PMF existed in the early universe. If confirmed by future observations, this PMF could be a new probe of physics before the CMB epoch and would impact our understanding of the physics of the early universe. 

\subsection{Model}
Recently, several different groups \cite{Lewis:2004ef,Yamazaki:2004vq,
Yamazaki:2006bq,Yamazaki:2006mi,
Yamazaki:2007oc,Giovannini:2008aa,Yamazaki:2008bb,Paoletti:2008ck,Caprini:2009vk,Kojima:2009ms} have been developing  the complex analysis of  PMF effects on the CMB and matter power spectrum.  
The work presented here, however, is the first complete statistical analysis of the constraints on PMF parameters.  
Although, occasional discrepancies  among different groups have occurred their sources have been identified and the groups are now concordant where comparisons can be made.  

To analyze the role of the PMF, one can assume that the photons and baryons behave as a single fluid since these particles interact rapidly before the last scattering of photons.  Briefly stated, the PMF directly affects ionized baryons through the Lorentz force. Since the baryons and photons are tightly coupled before the epoch of last scattering, photons are indirectly influenced by the PMF. The cold dark matter (CDM) is also indirectly affected by the PMF through gravitational interaction.
We assume that the PMF was generated  some time during the radiation-dominated epoch.  In the case of a flat Friedmann-Robertson-Walker (FRW) background cosmology with linear perturbations, the time evolution of the energy density of the PMF can be treated as a first order perturbation with a stiff source.  In this case, all of the back reactions from the fluid can be discarded.
For a PMF that is statistically homogeneous, isotropic and random, the fluctuation power spectrum can be parametrized\cite{Mack:2001gc} as a power-law
$S(k)= \langle B(k)B^\ast(k)\rangle  \propto k^{n_\mathrm{B}} $
where $n_\mathrm{B}$ is the power-law spectral index of the PMF that can be either negative or positive depending upon the
physical processes of magnetic field creation.
A two-point correlation function for the PMF can then be defined\cite{Mack:2001gc}  for $k < k_C$ where $k_C$  is a cutoff wave number in the magnetic power spectrum, i.e. 
$\left\langle B^{i}(\mbi{k}) {B^{j}}^*(\mbi{k}')\right\rangle 
	=	({(2\pi)^{n_\mathrm{B}+8}}/{2k_\lambda^{n_\mathrm{B}+3}})
		[{B^2_{\lambda}}/{\Gamma\left(\frac{n_\mathrm{B}+3}{2}\right)}]
		k^{n_\mathrm{B}}P^{ij}(k)\delta(\mbi{k}-\mbi{k}'), $
where $P^{ij}(k)=\delta^{ij}-\frac{k{}^{i}k{}^{j}}{k{}^2}~~.$
Here, $B_\lambda=|\mathbf{B}_\lambda|$ is the magnetic strength in comoving coordinates derived by smoothing over a Gaussian sphere of radius $\lambda=1$ Mpc.
The quantity  $k_\lambda = 2\pi/\lambda$.  Radiative viscosity damps the PMF \cite{Jedamzik:1996wp,Subramanian:1997gi}.   We use this effect to evolve the cutoff wave number, $k_\mathrm{C}$.
We then compute the PMF power spectrum using  the numerical methods developed in  
Refs.~\cite{Yamazaki:2006mi,Yamazaki:2007oc}.

In this article we place new limits on the  magnetic field strength together with other cosmological parameters using a Markov chain Monte Carlo (MCMC) method in the numerical analysis\cite{Lewis:2002ah}.
We use adiabatic initial conditions for the evolution of the CMB anisotropy and matter power spectrum in the presence of a PMF. Details of the initial conditions are summarized in Ref.~\cite{Lewis:2004ef,Kojima:2009ms}. 
For the scalar mode, we use adiabatic initial conditions for the matter contribution as in Ref. \cite{Kojima:2009ms}.
Although  in previous numerical estimations the curvature perturbations of the scalar mode were too small to stabilize the numerical calculations for larger scales at early times, our present initial conditions stabilize the  numerical scheme for all scales and times.  Hence, we can now obtain reliable numerical results\cite{footnote1}.

We consider  a flat  $\Lambda$CDM universe characterized by  8 parameters, i.e.~$\{ \Omega_b h^2,
	\Omega_c h^2,
	\tau_C,
	n_s,
	\log(10^{10}A_s),
	A_t/A_s$, $( |B_\lambda|/3.0\mathrm{(nG)})^4$, $n_\mathrm{B} \}$, 
where
$\Omega_b h^2$
and
$\Omega_c h^2$
are the baryon and CDM densities in units of the critical density,
$h$ denotes the Hubble parameter in units of 100 km s$^{-1}$Mpc$^{-1}$,
$\tau_C$
is the optical depth for Compton scattering,
$n_s$
is the spectral index of the primordial scalar fluctuations,
$A_s$
is the scalar amplitude of primordial scalar fluctuations and
$A_t$
is the scalar amplitude of the primordial tensor fluctuations.
We define the tensor index of the primordial tensor fluctuations as $n_t =-(A_s/A_t)/8 $.
For all cosmological parameters we use the same priors as those adopted in the WMAP analysis \cite{Dunkley:2008ie}. 

\subsection{Results}
In our likelihood analysis, the MCMC algorithm was performed until all of the cosmological parameters were well converged to the values listed in Table 1. With the inclusion of a PMF, the minimum total $\chi^2$  changes from 2803.4 to 2800.2 corresponding to a change in the $\chi^2$per degree of freedom from 1.033 to 1.031.  
Hence, the goodness of fit is slightly improved 
by considering a PMF even after allowing for new degrees of freedom.
Figure 1 shows the 68\% and 95\% C.L.~probability contours in the planes of
 various cosmological parameters versus the   amplitude $|B_\lambda|$, or power law index $n_\mathrm{B}$, along with the probability distributions.  The bottom of the figure shows the probability distributions for $|B_\lambda|$  and $n_\mathrm{B}$.  Of particular note for this article is the presence of  maxima for $|B_\lambda| = 0.85 \pm 1.25$ nG and    
$n_\mathrm{B}= -2.37^{+0.88}_{-0.73}$.  Although these values are consistent with zero magnetic field and thus only imply upper limits, they suggest the possibility that with forthcoming data (particularly for large CMB multipoles) a finite magnetic field may soon be detectable.

These figures exhibit no degeneracy between the PMF parameters and the standard  cosmological parameters. Table 1 confirms  that the standard cosmological parameters are not significantly different from those deduced directly from the WMAP 5yr data analysis without a PMF.
The reason  for this is  simple. The standard  cosmological parameters are mainly constrained by the observed CMB power spectrum for low multipoles $\ell < 1000$ (up to the 2nd acoustic peak).  On the other hand,   the PMF dominates for 
$\ell > 1000$.   Hence, the  PMF effect on the power spectrum is  nearly independent of the standard cosmological parameters.

The tensor to scalar ratio $A_t/A_s$ deduced in our analysis is smaller than the upper limit $A_t/A_s <$ 0.43 (95\% CL) deduced  from the WMAP 5yr data analysis without a PMF. 
The reason for this is that we define $A_t$ as the tensor amplitude of the primary CMB spectrum (without a PMF).  This tensor term only arises from the primordial gravitational-wave background produced during  inflation. We combine the tensor amplitude from the PMF and $A_t$ when we compare our tensor amplitude with the result deduced by others. The value of $A_t$ by itself is comparable  to the tensor contribution from the  PMF. Since the value of $A_t/A_s$ from the WMAP 5yr data is less than 0.43 (95\% CL),
our result,  is consistent with the previous constraints when the additional PMF contribution is included.

The  degeneracy of PMF parameters\cite{Yamazaki:2006bq} is broken by the different effects of the PMF on both the matter power spectrum and the CMB power spectrum.
The vector mode can dominate for higher $\ell$ in the CMB temperature anisotropies\cite{Yamazaki:2007oc}, while the matter power spectrum becomes sensitive\cite{Yamazaki:2006mi} to the power law spectral index $n_\mathrm{B}$ when a PMF is present.
Additionally, the PMF fluctuations are smaller  than the CMB fluctuations for the scalar mode on larger angular scales. Therefore,  the tensor to scalar ratio is not affected by the PMF. 

Figure 2 shows our deduced probability distributions and the 1$\sigma$ and $2\sigma$ (68\% and 95\% C.L.)
probability contours for the resultant cosmological parameters, $\sigma_8$, $H_0$, $z_\mathrm{reion}$, and Age.  Here, the "alternative normalization parameter" $\sigma_8$ is the root-mean-square of the matter density fluctuation within a comoving sphere of radius  $8h^{-1}$ Mpc. $H_0$ = 100$h$ km s$^{-1}$ Mpc$^{-1}$ is the Hubble parameter, $z_\mathrm{reion}$ is the red shift at which re-ionization occurs,  and Age is the presently observed age of the universe in Gyr.
It is important to keep in mind that these parameters are not  input parameters, but are  output results.  The concordance of the results on Figures 1 and 2 provides convincing evidence that both upper and lower limits to the parameters of the PMF can be deduced statistically.

Table 1 summarizes the upper limits to the PMF parameters along with input and output $\sigma_8$, $H_0$, $z_\mathrm{reion}$, and Age.  
In particular we find that 
$|B_\lambda| \mathbf{< 2.10} ( 68\% \mathrm{CL})$
nG and $\mathbf{< 2.98} ( 95\% \mathrm{CL})$ nG and  
$n_\mathrm{B} \mathbf{< -1.19} ( 68\% \mathrm{CL})$
and 
$\mathbf{< -0.25} ( 95\% \mathrm{CL})$ at a present scale of 1 Mpc.
Although previous work  
\cite{Yamazaki:2006bq,Yamazaki:2007oc,Yamazaki:2006mi,Yamazaki:2008bb}  could obtain a less stringent  upper limit to $|B_\lambda|$ they could not constrain $n_\mathrm{B}$ at all. Moreover, our deduced probability distributions suggest that a finite PMF provides the best fit.  

On angular scales smaller than that probed by the CMB the observed number density of galaxies is a better measure of the power spectrum.  Therefore, since the PMF mainly influences the  small angular scales, using both the CMB
(WMAP 5yr\cite{Hinshaw:2008kr},
ACBAR\cite{Kuo:2006ya},
CBI\cite{Sievers:2005gj},
Boomerang\cite{Jones:2005yb}) and the LSS (2dFDR\cite{Cole:2005sx}) observational data, 
 we can  constrain the PMF better than in previous works which relied on the CMB data only. 
In particular, the upper limit on $n_\mathrm{B}$ can be constrained for the first time and the lower limit to $n_\mathrm{B}$ is approaching the  1$\sigma$ confidence level. 

The right-bottom panel of FIG.~1 shows that the maximum likelihood is for $n_\mathrm{B}= -2.2^{+0.9}_{-1.3}$.
It is important\cite{Caprini:2001nb,Yamazaki:2006bq} to constrain  $n_\mathrm{B}$ as this parameter provides insight into models for the formation of the PMF.   If the PMF were formed during inflation one would expect a value of  
 $n_\mathrm{B} = -3$.  The value deduced in Fig.~1 and Table 1, are consistent with an inflation generated PMF at the 1$\sigma$ confidence level.  The optimum values  of $n_\mathrm{B}$, however, if confirmed would suggest that the PMF could have been generated at a later phase transition
\cite{Vachaspati:1991nm,Kibble:1995aa,Ahonen:1997wh,Joyce:1997uy}
(such as the electroweak transition) or even generated after the epoch of BBN \cite{Caprini:2001nb,Yamazaki:2006bq}.  This constraint on $n_\mathrm{B}$ derives from the gravity waves produced along with the PMF. Big-bang nucleosynthesis (BBN) depends upon a balance between the nuclear reaction rates and the expansion rate of the universe. Since gravity waves contribute to the total energy density they affect the expansion rate. Hence, they are constrained by a comparison between the BBN predictions and the  observed light element abundances \cite{Caprini:2001nb,Yamazaki:2006bq}.  
Forming a PMF at very late times (i.e. after BBN) is a challenge. However, models have been proposed (e.g.~via second-order MHD effects during recombination \cite{ichiki:2006sc}) for the generation of magnetic fields at late times which are not excluded from the range of PMF parameters of the 2 $\sigma$ confidence limits deduced here.

A PMF affects not only the temperature fluctuations, but also the polarization of the CMB.  
Although we fit all available polarization data, it turns out that the TT and BB modes (where T is the temperature fluctuation and B is the curl-like component of polarization) are most important.
  Figure 3 shows a comparison of the computed best-fit total power spectrum with the observed CMB spectrum.  Plots show various spectra for the TT and  BB modes.
We plot both the best fit and allowed regions both including  the SZ effect (scattering from re-ionized electrons) at the K(22.8GHz) band (upper curves) and without the SZ effect (lower curves).  Including the SZ effect only slightly diminishes the best fit magnitude of the PMF.
Although the CBI point falls about 1 $\sigma$ above the best fit, the $\chi^2$  is  dominated  by the better ACBAR08 data and this point does not significantly affect the deduced PMF parameters.. 

\subsection{Summary}
To summarize, this work presents the first complete statistical analysis of constraints on  parameters characterizing the PMF based upon available data on the CMB and matter power spectrum.  For the first time we constrain the spectral index $n_\mathrm{B}$, which provides new insight into models for the formation of the PMF. 
The matter power spectrum deduced here is consistent with observational constraints on the  growth of linear structure as is also reflected in the deduced  $\sigma_8$ parameter.
We conclude that future plans to observe the CMB anisotropies and polarizations for higher multipoles $\ell$, e.g.~via the {\it Planck}, {\it QUIET}, and {\it PolarBear} missions, we will be able to constrain the PMF more accurately.  This will permit a better understanding of the generation and evolution  of the PMF and provide new insight into the formation of  LSS as well as a possible new probe of the physics of the early universe.

\begin{acknowledgments}
This work has been supported in part by Grants-in-Aid for Scientific
Research (20105004, 20244035, 21740177) of the Ministry of Education, Culture, Sports,
Science and Technology of Japan.  This
work is also supported by the JSPS Core-to-Core Program, International
Research Network for Exotic Femto Systems (EFES).  Work at UND supported  by the US Department of Energy under research grant DE-FG02-95-ER40934.
\end{acknowledgments}

\begin{figure}
\includegraphics[width=0.8\textwidth]{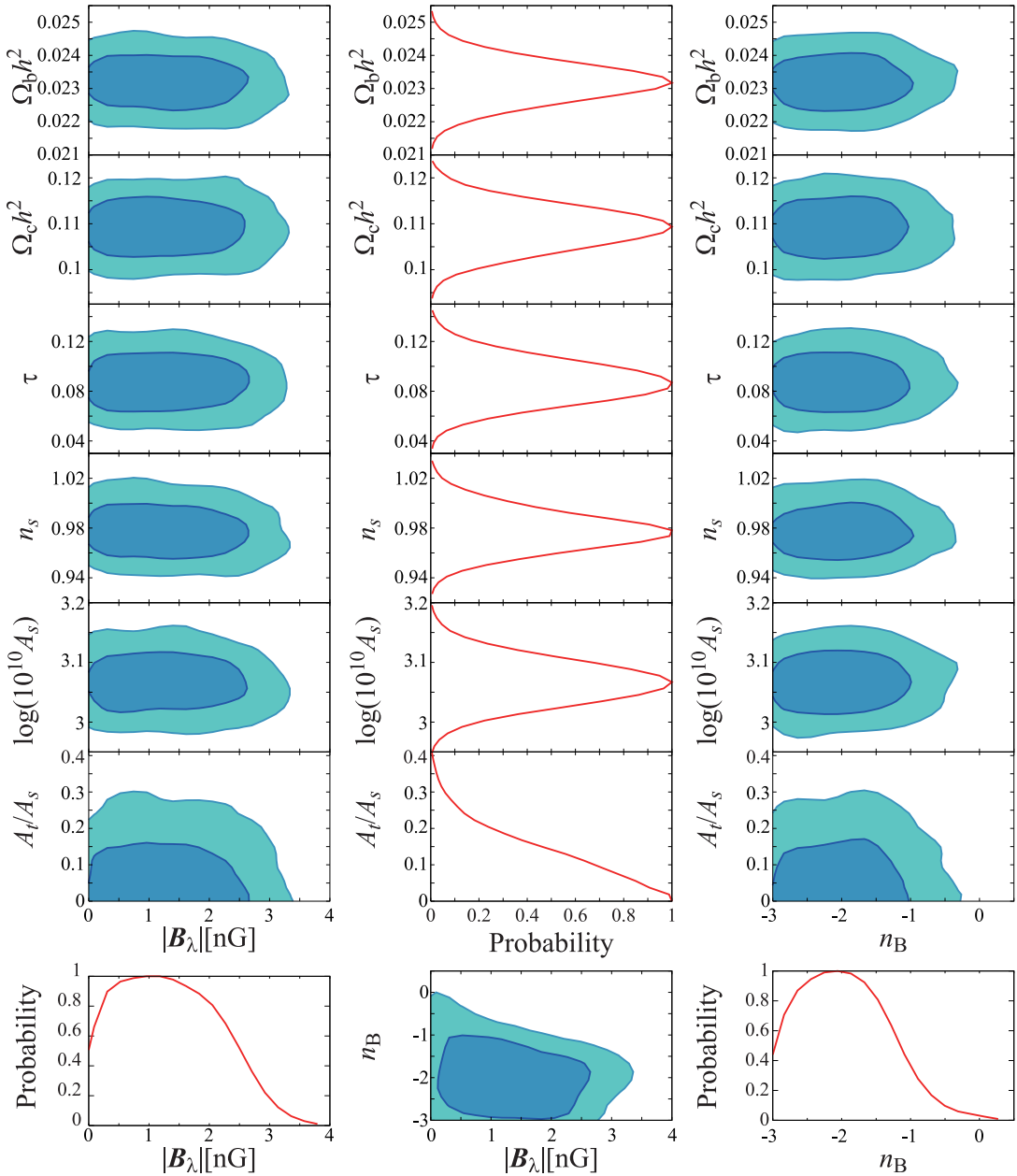}
\caption{\label{fig1} Probability distributions and contours of 1$\sigma$ and 2$\sigma$ confidence limits for the standard cosmological parameters as a function of the PMF field strength $|B_\lambda|$ and power law index $n_\mathrm{B}$.  
Blue contours show 1 $\sigma$(68\%)
 confidence limits and sky blue contours show 2 $\sigma$(95\%)
confidence limits. 
Red curves in the middle and bottom of the figure show the probability distributions of each parameter.  Note the existence of a maximum in the probability distributions for $|B_\lambda|$  and $n_\mathrm{B}$. 
The standard cosmological parameters do not have a degeneracy with the PMF parameters because they are mainly constrained by the observed CMB data for $\ell<$ 1000 (up to  the 2nd peak), while the PMF is mainly influenced by the power on smaller angular scales and higher multipoles, $\ell>$ 1000.
} 
\end{figure}

\begin{figure}
\includegraphics[width=1.0\textwidth]{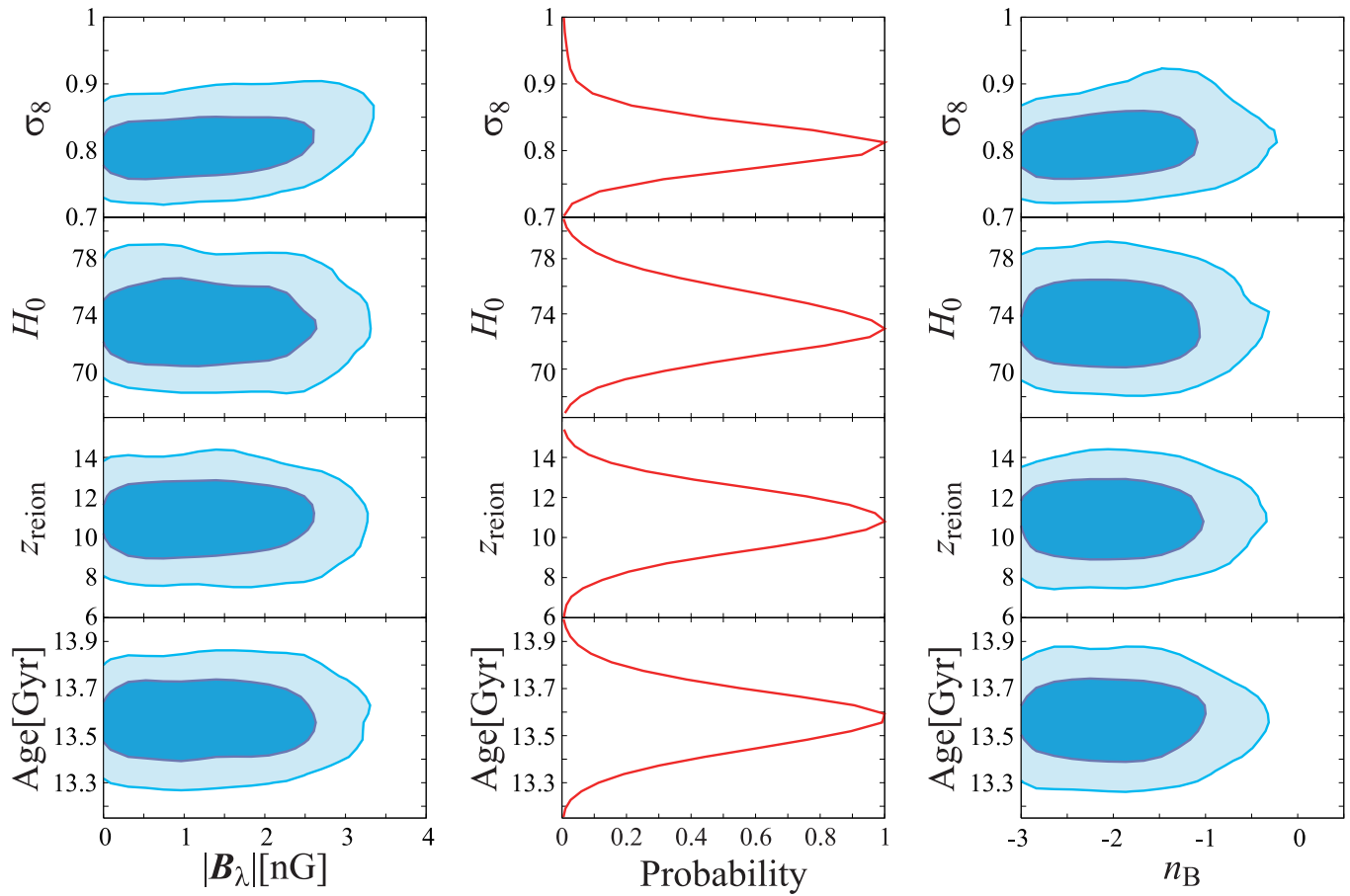}
\caption{\label{fig2} Probability distributions and contours of 1$\sigma$ and 2$\sigma$ confidence limits for the parameters  $\sigma_8$, $H_0$, $z_\mathrm{reion}$ and Age. 
Blue contours show 1 $\sigma$(68\%)
 confidence limits and sky blue contours show 2 $\sigma$(95\%)
confidence limits. 
Red curves show probability distributions for each parameter.
Note that these are not  input priors,  but output results.
} 
\end{figure}

\begin{figure}
\includegraphics[width=1.0\textwidth]{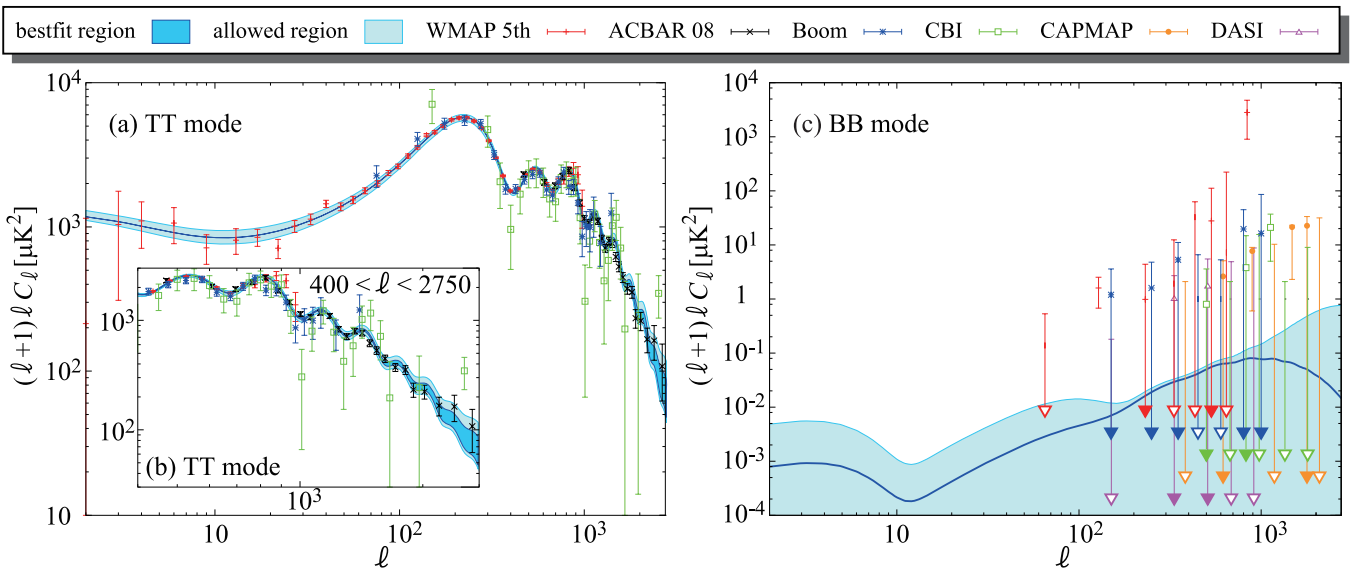}
\caption{\label{fig3}Comparison of the best-fit computed total power spectrum with the observed CMB spectra.  
Plots show various ranges for: (a) TT($2 < \ell < 2750$), (b) TT($400 < \ell < 2750$) and (c) BB($2 < \ell < 2750$) modes.
Blue regions are from the best-fit parameter set
and allowed regions are from the constrained parameter set in Table.1. 
Red, black, blue, green, orange, and purple dots with error bars show 
WMAP 5yr, ACBAR 08, Boomerang, CBI, CAPMAP, and DASI data, respectively. 
Downward arrows for the error bars on panel (c) indicate that the data points are upper limits.
Since the SZ effect depends upon the frequencies of the CMB for the TT mode, we plot the best fit and allowed regions in panels 3(a) and 3(b) surrounded by the curves with the SZ effect at the K(22.8GHz) band (upper curves) and without the SZ effect(lower curves).
Since the BB mode is not affected by the SZ effect, the upper curves and lower curves in  panel 3(c) are defined by the constrained cosmological parameters of Table.1.} 
\end{figure}

\begin{table}
\begin{tabular}{ccc} 
\multicolumn{3}{c}{\it Cosmological Parameters}\\
\hline
\multicolumn{1}{c}{Parameter} &
\multicolumn{1}{c}{
mean}&
\multicolumn{1}{c}{best fit}\\
\hline
$\Omega_b h^2$ &
$0.02320 \pm 0.00059$ &
$0.02295$ \\
$\Omega_c h^2$ &
$0.1094\pm0.0046$ &
$0.1093$ \\
$\tau_C$ &
$0.087\pm0.017$ &
$0.082$ \\
$n_s$ &
$0.977\pm0.016$ &
$0.970$ \\
$\ln(10^{10}A_s)$ &
$3.07\pm0.036$ &
$ 3.06$\\
$A_t/A_s$ &
$< 0.170 ( 68\% \mathrm{CL}), < 0.271 ( 95\% \mathrm{CL})$ &
$0.0088$\\
$|B_\lambda|\mathrm{(nG)}$ &
$\mathbf{< 2.10} ( 68\% \mathrm{CL}), \mathbf{< 2.98} ( 95\% \mathrm{CL})$ &
$\mathbf{0.85}$\\
$n_\mathrm{B}$ &
$\mathbf{< -1.19} ( 68\% \mathrm{CL}), \mathbf{< -0.25} ( 95\% \mathrm{CL})$ &
$\mathbf{-2.37}$\\
\hline
$\sigma_8$ &
$0.812^{+0.028}_{-0.033}$ &
$ 0.794$\\
$H_0$ &
$73.3\pm2.2$ &
$ 72.8$\\
$z_\mathrm{reion}$ &
$10.9\pm 1.4$ &
$ 10.5$\\
$\mathrm{Age(Gyr)}$ &
$13.57\pm 0.12$ &
$ 13.62$\\
\hline\hline
\end{tabular}
\caption{PMF parameters and $\Lambda$CDM model parameters and 68\%
confidence intervals ($A_t/A_s$  is a 95\% CL) from a fit to the WMAP\cite{Hinshaw:2008kr} + ACBAR\cite{Kuo:2006ya} + CBI\cite{Sievers:2005gj} +Boomerang\cite{Jones:2005yb} +  2dFDR\cite{Cole:2005sx} data. }
\end{table}

\bibliographystyle{apsrev}

\begin{thebibliography}{57}
\bibitem{footnote1}
Of course, the total magnetic influences on the  temperature fluctuations of the CMB are larger and more effective for higher $l$. Furthermore, the amplitude of the vector mode of the PMF for temperature fluctuations of the CMB is two orders of magnitude larger than that of the other  (scalar + tensor) modes for higher $l$. Thus, the contributions of the scalar mode from the PMF for constraining parameters (excluded $\sigma_8$ as output parameters) are of secondary importance.
\end{thebibliography}

\end{document}